\documentclass[12pt]{article}
\usepackage{amsmath,amsfonts,epsfig}

\advance \textheight by \topskip
\makeatletter
 \@addtoreset{equation}{section}
 \renewcommand\theequation
  {\ifnum\c@section>\z@\thesection.\fi
  \@arabic\c@equation}
\makeatother

\begin{document}

\title{
\begin{flushright}
{\small USACH-FM-00/11}\\[1.0cm]
\end{flushright}
{\bf
Nonlinear Supersymmetry, Quantum Anomaly
and Quasi-Exactly Solvable Systems}}

\author{{\sf Sergey M. Klishevich${}^{a,b}$}\thanks{E-
mail:
sklishev@lauca.usach.cl}
{\sf and Mikhail S. Plyushchay${}^{a,b}$}\thanks{E-
mail: mplyushc@lauca.usach.cl}
\\
{\small {\it ${}^a$Departamento de F\'{\i}sica,
Universidad de Santiago de Chile,
Casilla 307, Santiago 2, Chile}}\\
{\small {\it ${}^b$Institute for High Energy Physics,
Protvino, Russia}}}
\date{}

\maketitle

\vskip-1.0cm

\begin{abstract}
The nonlinear supersymmetry of one-dimensional  systems
is investigated in the context of the quantum anomaly
problem. Any classical supersymmetric system
characterized by the nonlinear in the Hamiltonian
superalgebra  is symplectomorphic to a supersymmetric
canonical system with the holomorphic form of the
supercharges. Depending on the behaviour of the
superpotential,  the  canonical supersymmetric systems are
separated into the three classes. In  one of them the
parameter specifying  the supersymmetry order is subject to
some sort of classical quantization, whereas the
supersymmetry of another extreme class has a rather fictive
nature since its fermion degrees of freedom are decoupled
completely by a canonical transformation. The nonlinear
supersymmetry with polynomial in momentum supercharges
is analysed,  and  the most general one-parametric
Calogero-like solution with the second order supercharges
is found. Quantization of the systems of the canonical form
reveals the two anomaly-free classes, one of which gives
rise  naturally to the quasi-exactly solvable systems. The
quantum anomaly problem for the Calogero-like models  is
``cured'' by the specific superpotential-dependent term of
order  $\hbar^2$. The nonlinear  supersymmetry admits the
generalization to the case of two-dimensional  systems.
\end{abstract}

\section{Introduction}

After introducing the supersymmetric quantum mechanics
as a toy model for studying the supersymmetry
breaking mechanism \cite{wit,wit1}, it was applied  for
solving
many problems in  theoretical and mathematical physics
\cite{susyqm, cooper}.
The  most recent applications of the supersymmetric
quantum mechanics can be found
in the dynamics of D-branes and black holes
\cite{BhDb,BhDb1},
in M-theory and matrix
models
\cite{matrix}, in the theory of integrable systems
\cite{cooper}
and fluid mechanics \cite{fluid,fluid1}.

Some time ago  it was observed that the
pure parabosonic \cite{susy-pb}
(and  parafermionic \cite{susy-pf})  systems possess
the supersymmetry characterized by the  nonlinear
superalgebra. Such  nonlinear supersymmetry
takes place in the  simple quantum mechanical system
generalizing the usual superoscillator \cite{susy-pb},
and  earlier it was revealed in a similar but particular  form
in the fermion-monopole system~\cite{machol,sfm}, and
in the  $P,T$-invariant systems of planar
fermions \cite{ptf} and  Chern-Simons fields \cite{ptch}.
The algebraic structure of the nonlinear supersymmetry
resembles the structure  of  the finite
$W$-algebras \cite{Walg}
for which the commutator of any two generating elements
is proportional to a finite order polynomial in them.

As it was noted in Ref. \cite{susy-pb}, in a generic case
under  attempt of constructing  the quantum analogue of the
classical systems possessing the  nonlinear supersymmetry
one faces the problem of the quantum anomaly
\cite{qa1,qa2,qa3}.
To resolve this problem, in the present  paper we
investigate the nonlinear
supersymmetry of one-dimensional systems
at the  classical and quantum levels.
This will allow us to reveal the
unexpected very close relation of the nonlinear
supersymmetry with associated quantum anomaly problem
to the quasy-exactly solvable systems
\cite{turbiner,shifman,ushv,olv,dunne} being related,
in turn, to the conformal field theory
\cite{morozov,gor,bazh,tateo,suzuki}.
The results will give also a new perspective
on the known quantum supersymmetry
characterized by the second order supercharges
\cite{andrian}.

The paper is organized as follows.
Section \ref{classics} is devoted to the detailed
investigation of the various  aspects of the classical
one-dimensional systems possessing the nonlinear
supersymmetry of the most general form.
In Section 3 we consider  their  quantization,
and reveal the two classes of
anomaly-free quantum systems possessing the
nonlinear supersymmetry of arbitrary order
$k\in \mathbb Z_+$.
One of them turns out to be  closely related to
the quasy-exactly solvable systems
and this aspect of the nonlinear supersymmetry is
investigated in Section 4.
Section 5 discusses the general quantum case of the
$k=2$ supersymmetry in the context of anomaly-free
quantization of the class of $k=2$ Calogero-like
supersymmetric systems found and analyzed in Section 2.
We show also how this $k=2$ supersymmetry
can be used for constructing new exactly solvable systems.
In Section 6 the brief summary of the obtained results
is presented and some open problems to be interesting
for further investigation are discussed.
In particular, we point out how  the nonlinear
supersymmetry can be generalized to the case of
the two-dimensional classical and quantum systems.

\section{Classical supersymmetry}
\label{classics}

In this section we  investigate
the classical supersymmetry of the most general form
realizable in one-dimensional  boson-fermion system.
We shall  show that in generic case the supersymmetry
is characterized by  a nonlinear Poisson algebra and
includes the usual supersymmetry as a particular case.
Analysing the structure of the supersymmetry
from the viewpoint of canonical transformations,
we shall observe the existence of  the three essentially
different classes: in the first class  the parameter
characterizing  the order of superalgebra  is subject to
the classical quantization, in the second (intermediate)
class the supercharges' Poisson bracket can be equal to
any real nonnegative degree of the Hamiltonian,
whereas the systems of the third class allow ones the
complete classical decoupling of the fermion from the boson
degrees of freedom.

\subsection{General structure of supersymmetry}
\label{c1}

Following Ref. \cite{clas}, let us consider a
non-relativistic particle in one dimension
described by the Lagrangian
\begin{equation}\label{Lgen}
 L=\frac 12\dot x^2-V(x)-L(x)N+i\theta^+\dot\theta^-,
\end{equation}
where $\theta^\pm$ are the Grassmann variables,
$(\theta^+)^*=\theta^-$, $N=\theta^+\theta^-$, and
$V(x)$ and $L(x)$ are two real functions. The
nontrivial Poisson-Dirac brackets for the system are
$\{x,\,p\}=1$ and $\{\theta^+,\,\theta^-\}=-i$, and the
Hamiltonian is
\begin{equation}\label{Hgen}
 H=\frac 12p^2+V(x)+L(x)N.
\end{equation}
The latter generates the equations of motion
\[
 \dot{x}=p,\qquad \dot{p}={}-V'(x)-L'(x)N,\qquad
 \dot{\theta}^{\pm}={}\pm iL(x)\theta^{\pm}.
\]
The Hamiltonian $H$ and the nilpotent quantity $N$ are
the
even integrals of motion for any choice of the
functions $V(x)$, $L(x)$, whereas the odd quantities
\begin{equation}
\label{Qd}
 Q^\pm=B^\mp(x,\,p)\theta^\pm, \quad
 (B^+)^*=B^-,
\end{equation}
are the integrals of motion when
the differential equations
\begin{equation}
\label{tQ=0}
 \left(p\frac\partial{\partial x}-V'(x)\frac\partial
 {\partial p}\mp iL(x)\right)B^\pm(x,\,p)=0
\end{equation}
have the solutions being regular
functions in the corresponding domain of the phase
space defined by $V(x)$ and $L(x)$. It is obvious that
such odd integrals can exist only for a special choice
of the functions $V(x)$ and $L(x)$. Let us investigate
this question in detail and restrict ourselves to the
physically interesting class of the systems given by
the potential $V(x)$ bounded from below. Such a
potential can generally be represented in terms of a
superpotential $W(x)$ and real constant $v$:
\[
 V(x)=\frac 12W^2(x)+v.
\]
The condition of regularity of $B^\pm(x,p)$ at $p=0$
leads to the relation $L(x)=W'(x)\phi(x)$ with some
function $\phi(x)$. Having in mind that for the
functions $B^\pm(x,\,p)$ the substitution $p\to-p$ is
equivalent to the complex conjugation, one can
represent them as $B^\pm(x,\,p)=B(W(x),\,\mp ip)$. Then
the Hamiltonian and Eq. (\ref{tQ=0}) take the form
\begin{equation}\label{HWgen}
 H=\frac 12p^2+\frac 12W^2(x)+v+W'(x)\tilde \phi(W)N,
\end{equation}
\begin{equation}
 \left(p\frac{\partial}{\partial W}-W\frac{\partial}
 {\partial p}+ i\tilde{\phi}(W)\right) B\left( W,\,ip
 \right) =0, \label{BeqW}
\end{equation}
where $\tilde{\phi}(W(x))=\phi(x)$. In terms of the
complex variables
\begin{equation}
 z=W(x)+ip,\qquad \bar{z}=W(x)-ip
 \label{b}
\end{equation}
Eq. (\ref{BeqW}) is represented as
\begin{equation}
 \left(\bar z\partial _{\bar z}-z\partial _z
 +\tilde\phi\left(\tfrac{z+\bar z}2\right)\right)
 B(z,\,\bar{z})=0.
 \label{ez}
\end{equation}
General solution to Eq. (\ref{ez}) is given by
\begin{equation}\label{Bgs}
 B(\rho,\varphi)=f(\rho)\exp\left(i\int_{\varphi _0}^
 \varphi\tilde\phi(\rho\cos\lambda)\,d\lambda\right),
\end{equation}
where $f(\rho)$ is an arbitrary function and
$z=\rho e^{i\varphi}$. The simplest case
$\phi(x)=\alpha\geq 0$ with $f(\rho)=\rho^\alpha$
corresponds to the holomorphic solution of
Eq. (\ref{ez}), $B(z)=z^\alpha$, whereas the case
$\alpha\leq 0$ with $f(\rho)=\rho^{-\alpha}$ gives the
antiholomorphic solution $B(\bar z)=\bar z^{-\alpha}$,
both regular at $z=\bar z=0$. If the superpotential
$W(x)$ is the unbounded function, one of the functions
$B(z)=z^\alpha$ or $B(\bar{z})=\bar{z}^{-\alpha}$ is
well defined on the whole complex plane only for
$\alpha\in \mathbb Z$, i.e. we have here some sort of
classical quantization  \cite{clasq}. This simplest
solution with $ \alpha\in \mathbb Z$ corresponds to the
nonlinear supersymmetry investigated in
Ref.~\cite{susy-pb}, and  includes the usual linear
supersymmetry with $\vert \alpha\vert =1$ as a
particular case.

According to the definition (\ref{Qd}), the functions
$B^\pm(x,\,p)$ are defined up to an additive nilpotent
term proportional to\footnote{At the quantum level the
terms
$N\theta^\pm$, $N^2$ etc. do not vanish
and may be essential for preserving the
supersymmetry.}
$N$. This allows ones to represent
the supercharges in the equivalent
form
\begin{equation}\label{Qgen}
 Q^\pm=f(H)e^{\pm i\Phi(p,W(x))}\theta^\pm,
 \quad\Phi(p,\,W(x))=
 \int _{\varphi _0}^\varphi\tilde\phi(\rho\cos\lambda)
 \,
 d\lambda .
\end{equation}
We suppose here that like in the case $\phi(x)=\alpha$,
the function $f(H)$ is chosen in the simplest form
compatible with the requirement of regularity of the
supercharges (see below).
The supercharges (\ref{Qgen}), the Hamiltonian
(\ref{HWgen}) and the nilpotent integral
$N$ form
generally the nonlinear superalgebra
\[
 \left\{Q^-,\,Q^+\right\}=-if^2(H),
 \quad
 \{N,Q^\pm\}=\mp iQ^\pm,
 \quad
 \{Q^\pm,H\}=0,\quad
 \{N,H\}=0.
\]

\subsection{Three cases of supersymmetry and classical
quantization}
\label{c2}

In the  case of unbounded superpotential $W(x)$, the
dynamics of the system projected on the unity of the
Grassmann algebra is defined on the whole complex plane
$\mathbb C$. In order to have the supercharges
(\ref{Qgen}) to be well defined single-valued
observables on $\mathbb C$, we have to impose the
condition
\[
 \int _0^{2\pi}\tilde\phi\left(\rho\cos\lambda\right)
 \,d\lambda=2\pi k,\qquad k\in\mathbb Z.
\]
The function  $\tilde\phi$ is supposed to be
regular and  can be represented in the form
\begin{equation}
 \tilde\phi(W(x))=k+W(x)M(W^2(x)),
 \label{tphi}
\end{equation}
where $M(W^2)$
is an arbitrary function, $\vert M(0)\vert <\infty$.
As a consequence, the supercharge can be written as
\begin{equation}\label{QkZ}
 Q^+=z^ke^{i\int
 _0^pM(p^2-y^2+W^2(x))\,dy}\theta^+.
\end{equation}
Here we suppose that $k$ is nonnegative; in the case of
negative $k$ equation (\ref{QkZ}) with substitutions
$k\rightarrow -k$,
$\theta^+\rightarrow \theta^-$ gives
the supercharge $Q^-=(Q^+)^*$.
The exponential factor in Eq. (\ref{QkZ}) can be
removed
by the transformation
\begin{equation}\label{CT}
 \Theta^\pm=e^{\pm iG(p,\,x)}\theta^\pm,\qquad
 X=x+\partial _pG(p,x)N,\qquad
 P=p-\partial _xG(p,x)N,
\end{equation}
that is the canonical transformation
with generating function $G$ obeying the
differentiability condition
\[
 \partial _p\partial _xG(p,x)=
 \partial _x\partial _pG(p,x).
\]

In the case of general superpotential $W(x)$,
the transformation (\ref{CT}) can be used to reduce the
classical Hamiltonian
(\ref{HWgen}) to the most simple form, e.g., to the
form
with $L(x)=\alpha W'(x)$, $\alpha=const$.
This gives the following equation for the function
$G$:
\begin{equation}
\label{GeqW}
 \left(p\frac{\partial}{\partial x}-W(x)W'(x)
 \frac{\partial}{\partial p}\right)G(x,\,p)
 +W'(x)\left(\tilde\phi(W(x))-\alpha\right)=0.
\end{equation}
Though the equation (\ref{GeqW}) has exactly the form
of that for $\log B$ with shifted $\tilde\phi$ (see Eq.
(\ref{BeqW})), there is a difference: we require for
$G$ to be regular and single-valued function on the
whole physical domain, whereas the same condition of
regularity is imposed on the function $B$, but not on
$\log B$. The general solution to Eq. (\ref{GeqW}) is
\[
 G(p,\,x)=\int _{\varphi _0}^\varphi\left(\tilde\phi
 (\rho\cos\lambda)-\alpha\right)\,d\lambda ,
\]
and its behaviour depends on the physical domain for
$z$ defined, in turn, by the properties of the
superpotential $W(x)$. Here it is necessary to separate
the three different cases and the results can be
summarized as follows.

\begin{enumerate}
\item The physical domain in terms of $z$ includes the
origin ($a<W(x)<b$, $a<0$,
$b>0$).  This, in particular, corresponds to the
case of unbounded superpotential with
$a=-\infty$, $b=+\infty$.
{}From the regularity of $G$
in such a domain it follows that $\alpha=\tilde
\phi(0)$. We assume that the function $\tilde\phi(W)$
can be decomposed into the
Taylor series at $W=0$. {}From the regularity of $B$
we arrive at the classical ``quantization''
condition $\alpha=k$, $k\in\mathbb Z$, and at
the restriction (\ref{tphi}) on the function
$\tilde\phi(W)$. Thus, the most general form of the
Hamiltonian admitting the nonlinear supersymmetry is
\begin{equation}
\label{h1}
 H=\frac{p^2}2+\frac 12W^2(x) + v
 +W'(x)\left[k+W(x)M(W^2(x))\right]N,\qquad k\in
 \mathbb
 Z,
\end{equation}
whose associated supercharges have been described
above. By the canonical transformation (\ref{CT}) with
$G(p,\,x)=\int _0^p M(p^2-y^2+W^2(x))\,dy$
(supplemented by the transformation
$\Theta^\pm\to\Theta^\mp$ in the case $k<0$) we can
always reduce the system with this Hamiltonian and the
supercharge (\ref{QkZ}) to the form of the
supersymmetric system possessing the holomorphic
supercharge:
\begin{equation}\label{hg}
 H=\frac 12P^2+\frac 12W^2(X)+v
 +kW'(X)\Theta^+\Theta^-,\qquad
 Q^+=Z^k\Theta^+, \qquad k\in\mathbb Z_+,
\end{equation}
where $Z=W(X)+iP$.
The presence of the ``quantized'', integer number $k$
in the Hamiltonian (\ref{hg}) means that the instant
frequencies of the oscillator-like odd, $\Theta^\pm$,
and even, $Z$, $\bar{Z}$, variables are commensurable.
Only in this case the regular odd integrals of motion
can be constructed, and the  factor $Z^k$
in the supercharge $Q^+$ corresponds to the $k$-fold
conformal mapping of the complex plane (or the strip $a
<\rm{Re}\, Z<b$) on itself (or on the corresponding
region in $\mathbb C$).

\item The physical domain is defined by the condition
$\mathop{\mathrm{Re}}z\geq 0$ (or $\mathop{\mathrm{
Re}}
z\leq 0$) and also includes the origin of the complex
plane. But unlike the previous case, there are no
closed contours around $z=0$. As a consequence, though
the regularity of $G$ results in the same relation
$\alpha=\tilde\phi(0)$, no ``quantization'' condition
appears from the regularity of $B$. The most general
form of the Hamiltonian and the supercharge is
\begin{equation}\label{h2}
 H=\frac{p^2}2+\frac 12W^2(x)+v
 +W'(x)\left[\alpha+R(W(x))\right]N,\quad
 Q^+=z^\alpha e^{i\int _{\varphi _0}^\varphi
 R(\rho\cos\lambda)\,d\lambda}\theta^+,
\end{equation}
where we assume that $\alpha\in\mathbb R$, and the
function $R(W)$ is analytical at $W=0$ and
\mbox{$R(0)=0$}. The singularity in $Q^+$ at
the origin $z=0$ for $\alpha<0$ is not physical and can
be removed multiplying $Q^+$ by
$(\bar zz)^{-\alpha}$, that results in changing the
holomorphic function  $B(z,\bar z)=z^\alpha$   for
the antiholomorphic function
$B(z,\bar z)=\bar z^{-\alpha}$  to be regular at
$\bar{z}=0$.
After the canonical
transformation (\ref{CT}) with the function $G(p,\,x)=
\int_{\varphi_0}^\varphi R(\rho\cos\lambda)\,d\lambda$
(supplemented by the transformation
$\Theta^\pm\to\Theta^\mp$ for $\alpha<0$), the
Hamiltonian and the supercharge can be reduced
to the form
\begin{equation}\label{h3}
 H=\frac 12P^2+\frac 12W^2(X)+v+
 \alpha W'(X)\Theta^+\Theta^-, \quad
 Q^+=Z^\alpha\Theta^+,\quad\alpha\in\mathbb R_+.
\end{equation}

\item The physical domain is defined by the condition
$\mathop\mathrm{Re}z>0$ (or $\mathop{\rm Re}z<0$),
i.e.
the origin of
the complex plane is not included. In this case
$\alpha$ and $\tilde\phi(0)$ are not related since the
function $G$ admits in such a domain the terms
proportional to $\arg z$. Therefore, though the general
form of the Hamiltonian and the supercharge is
\begin{equation}\label{h4}
 H=\frac{p^2}2+\frac 12W^2+v+W'\tilde\phi(W)N,
 \qquad
 Q^+=f(H)e^{i\int _{\varphi _0}^\varphi\tilde
 \phi(\rho\cos\lambda)\,d\lambda}\theta^+ ,
\end{equation}
by the canonical transformations (\ref{CT}) with the
function $G(p,\,x)=\int_{\varphi_0}^\varphi\tilde
\phi(\rho\cos\lambda)\,d\lambda$, one can reduce the
Hamiltonian to the form
\begin{equation}\label{decouple}
 H=\frac 12 P^2+\frac 12 W^2(X)+v
\end{equation}
with trivial dynamics for the Grassmann variables
$\Theta^\pm$, $\dot\Theta^\pm=0$, playing the role of
the supercharges. Thus, {\it classically} the
supersymmetry of any system with bounded non-vanishing
superpotential has a ``fictive'' nature.
\end{enumerate}

The obtained classification of the classical
supersymmetric
systems emerged from the aim to present
the Hamiltonian in the most simple form when
the superpotential $W(x)$ has the definite behaviour
defining  the type of the
physical domain in the complex plane.
On the other hand, one can consider the supersymmetry
from the viewpoint of the functional dependence of the
Hamiltonian on $W(x)$ without specifying  the
superpotential's type.  Then the Hamiltonian and the
supercharges
\begin{equation}
\label{hg0}
 H=\frac 12p^2+\frac 12W^2(x)+v
 +kW'(x)\theta^+\theta^-,\quad
 Q^+=z^k\theta^+, \quad
 Q^-=\bar{z}{}^k\theta^-, \quad
 k\in\mathbb Z_+
\end{equation}
give the intersection of the described three classes of
the systems,
and can be treated as the representatives of the
more broad classes of the supersymmetric systems
(\ref{h1}), (\ref{h2}) and
(\ref{h4}),
with which (\ref{hg0})
is  related by the corresponding
symplectomorphism.
For  (\ref{hg0}) the constant $k$ characterizes the
degree
of nonlinearity of the associated superalgebra
and one can refer to it as to the system with
$k$-supersymmetry.
At the same time, it is necessary to bare in mind that
in the
case of the superpotential of the third  class  all the
systems
(\ref{hg0}) with different $k\neq 0$ are symplectomorphic
to the system with $k=0$.

The three types of supersymmetric
Hamiltonians (\ref{h1}), (\ref{h2}), (\ref{h3}) are
defined up to the additive constant $v$. This
arbitrariness could be used to present the potential
$V(x)$ in terms of other superpotential via the
relation
\begin{equation}\label{trans}
 \tilde W^2(x)=W^2(x)+const.
\end{equation}
Generally, the superpotentials $W(x)$ and $\tilde{W}(x)
$ can correspond to different types of nonlinear
supersymmetry. Then the natural question is:
can such a transition change the type of supersymmetry
when the form of the nilpotent term of the Hamiltonian
has been already fixed? In other words, the question
is if the described classification of {\it classical}
supersymmetric Hamiltonians has an invariant sense. The
invariance of the classification is demonstrated  in
Appendix. At the same time, it is necessary to stress
that the systems related by the canonical
transformation have not to be equivalent on the quantum
level due to the ordering problem and because the
canonical transformations are, as a rule,
nonpolynomial in momenta.

For the sake of completness, let us discuss the
Lagrangian formulation for the $k$-supersymmetric
system
(\ref{hg0}).
Its  Lagrangian is
\begin{equation}
\label{l1}
 L=\frac 12\dot x^2-\frac 12W^2(x)-v-kW'(x)\theta^+
 \theta^-+i\theta ^+\dot\theta^-.
\end{equation}
In the Hamiltonian formulation the
supertransformations of
the variables $x$ and $\theta^\pm$ are
generated canonically by the supercharges:
\begin{align*}
 \delta x&=\{x,\,Q^+\}\eta^--\{x,\,Q^-\}\eta^+
 ={ik}
\left[z^{k-1}
\theta^+\eta^-+\bar{z}^{k-1}
\theta^-\eta^+\right],\\
 \delta\theta^+&=\{\theta^+,\,Q^-\}\eta^+
 ={}- iz^k\eta^+,\quad
 \delta\theta^-
 =-\{\theta^-,\,Q^+\}\eta^-
 ={} i\bar{z}{}^k\eta^-.
\end{align*}
Using the equations of the motion, we
obtain the corresponding supertransformations at the
Lagrangian level:
$$
 \delta x={ik}\left[\left(A^-\right)^{k-1}
 \theta^+\eta^-+\left(A^+\right)^{k-1}
 \theta^-\eta^+\right],
 \qquad
 \delta\theta^\pm={}\mp i\left(A^\pm\right)^k\eta^\pm,
$$
where $A^\pm=W(x)\mp i\dot x$. The
Lagrangian (\ref{l1}) is quasi-invariant under
these supertransformations:
$$
\delta L=\frac
d{dt}\left[i k\dot x\Bigl((A^-)^{k-1}
\theta^+\eta^-+(A^+)^{k-1} \theta^-\eta^+\Bigr)\right].
$$
It is worth noting that on shell the commutator
of the two supertransformations for the physical variables
 is proportional to the translation in
time:
$$
[\delta_1,\delta_2]{\cal X}=-i2^k k E^{k-1}(x,\dot x,
\theta^\pm)(\eta^-_1 \eta^+_2-\eta^-_2\eta^+_1)
\frac d{dt}{\cal X},
$$
where ${\cal X}$ is $x$ or $\theta^\pm$, and
$E(x,\dot x,\theta^\pm)$ is the energy function of the
system. Therefore, the supertransformations form an
open algebra with the structure functions depending on
the physical variables.

\subsection{Supersymmetry with polynomial supercharges}
\label{PolynomCase}

We have arrived at the three types of Hamiltonians
(\ref{h1}), (\ref{h2}), (\ref{h3}), which after
appropriate canonical transformations can be reduced to
the more simple form with the associated supercharges
represented in the holomorphic or antiholomorphic form.
On the other hand, as it was mentioned above, the
quantization of canonically equivalent classical
systems can give in some cases the quantum systems with
different types of supersymmetry, even of different
order $k$. Therefore, the search for other special
representations for the Hamiltonian and associated
supercharges is important. Based on this remark, let us
look for the representation in which the function
$B(x,\,ip)$ defining the supercharges is the polynomial
in $p$ of the degree $k$, i.e.
\[
 B(x,\,ip)=\sum_{n=0}^kb_{k-n}(x)(ip)^n.
\]
Substituting this into Eq. (\ref{tQ=0}),
we obtain the recurrent equation
\begin{equation}
 b_n'(x)+\left(k-n+2\right)b_{n-2}(x)V'(x)-L(x)
 b_{n-1}(x)=0, \label{bn}
\end{equation}
where $b_n(x)=0$ for $n<0$ and $n>k$ is assumed.
Due to the equation $b_0'(x)=0$, one can fix
 $ b_0(x)=1$. Then the part of the equations
(\ref{bn}) can be solved giving for
$L(x)$ and $V(x)$
the relations
$$
 L(x)=b_1'(x), \qquad
 V(x)=\frac 1k\left(\frac{b_1(x)^2}2-b_2(x)\right)+v.
$$
To simplify the notation,
we put $b_1(x)=y(x)$ and
realize the change of the variables
\begin{equation}\label{x(y)}
 x=x(y) ,\qquad \varphi _n(y)=b_n(x(y)),
 \quad\text{for} \quad n>1.
\end{equation}
If the function inverse to $y(x)$ does not exist
globally,
we can perform this transformation separately on
each interval where the function $y(x)$ is monotonic.
Under the transformation (\ref{x(y)}), Eq. (\ref{bn})
acquires the form of the system of differential
equations in
the variable $y$:
\begin{equation} \label{phn}
 \varphi _n'(y)-\frac{k-n+2}{k}\varphi _{n-2}(y) \left(
 \varphi _2'(y)-y\right)-\varphi _{n-1}(y) =0.
\end{equation}
A general solution to this system has
$k-1$ real parameters. If we know solution to the
system (\ref{phn}), we could find the form of the
functions $b_n(x)$, at least implicitly.
This means that the general solution to
Eq. (\ref{bn}) depends on arbitrary function $b_1(x)$
but not on its derivatives, and
which can be called the superpotential.
One notes also that the system (\ref{phn})
is invariant under the transformation $y\to -y$ if the
functions $\varphi _n$ obey the relation
\begin{equation}
 \varphi _n\left( -y\right) =\left( -1\right) ^n
 \varphi _n \left( y \right) . \label{parity}
\end{equation}
The solution corresponding to the
holomorphic case is of this type.

The simplest case $k=1$ corresponds to the usual
linear supersymmetry, and we turn to the case
$k=2$. For $k=2$,
from (\ref{phn}) we obtain the equation for $\varphi _2
$:
\[
 y\varphi _2'+2\varphi _2=y^2.
\]
This equation has the solution
\[
 \varphi _2=\frac{y^2}4+\frac{4c}{y^2},
\]
where $c$ is an arbitrary real constant. Considering
$W(x)= y(x)/2$ as a superpotential, one arrives at the
supercharges of the form
\begin{equation}\label{CalogeroQ}
 Q^\pm=\frac 12\left[\left({}\pm ip+W(x)\right)^2
 +\frac{c}{W^2(x)}\right]\theta^\pm,
\end{equation}
which together with the Hamiltonian
\begin{equation} \label{CalogeroH}
 H=\frac 12\left[p^2+W^2(x)-\frac{c}{W^2(x)}\right]
 +2W'(x)N+v
\end{equation}
form the nonlinear superalgebra
\begin{equation}
\label{qqh}
 \left\{Q^-,\,Q^+\right\}={}-i\left(\left(H-v\right)^2
 + c \right),
 \qquad
 \{Q^\pm,H\}=0.
\end{equation}
Note that the Hamiltonian
(\ref{CalogeroH}) has the Calogero-like form: at
$W(x)=x$ its projection to the unit of Grassmann
algebra takes the form of the Hamiltonian of the
two-particle Calogero system.
The functions $B^\pm(p,\,x)=B(x,\mp ip)$ and
Hamiltonian (\ref{CalogeroH}) form the nonlinear
Poisson algebra
\[
 \left\{B^\pm,\,H\right\}=\pm iW'B^\pm,\qquad
 \left\{B^-,\,B^+\right\}=-4iW'H,
\]
which does not depend on the constant $c$,
and is reduced to the $sl(2,R)$ algebra at
$W(x)=x$.

For $c=0$ the obtained $k=2$ supersymmetric system
(\ref{CalogeroQ}), (\ref{CalogeroH}) is reduced to the
$k=2$ supersymmetric system of the form (\ref{hg0})
characterized by the holomorphic form of the
supercharge.  Let us investigate the relationship of
the
system (\ref{CalogeroQ}), (\ref{CalogeroH})
with $k$-supersymmetry (\ref{hg0}) in the case
$c\ne 0$.
First we  show that for $c<0$ ($c=-\gamma^2$,
$\gamma>0$) the system (\ref{CalogeroH}),
(\ref{CalogeroQ}) can be reduced to the linear ($k=1$)
supersymmetric system of the form (\ref{hg0}) with the
corresponding holomorphic supercharge multiplied by
some function of the Hamiltonian. To show this we
represent the potential of the Hamiltonian
(\ref{CalogeroH}) in the form
\[
 V(x)=\frac 12\left(W(x)-\frac\gamma{W(x)}\right)^2
 +v+\gamma=\frac 12\tilde W^2(x)+v+\gamma ,
\]
Without loss of generality we can consider $W(x)>0$ and
get the relation
\[
 W(x)=\frac 12\left(\tilde W(x)+
 \sqrt{\tilde W^2(x)+4\gamma}\right).
\]
Therefore, in terms of $\tilde W(x)$
the function $L(x)$ is represented as
\[
 L(x)=\tilde W'(x)\left[1+\tilde W(x)M(\tilde W{}^2(x))
 \right],
 \quad
 M(\tilde W{}^2(x))=
 \left(
 \tilde W^2(x)+4\gamma\right)^{-1/2}.
\]
Comparing this with the factor at the nilpotent term in
Hamiltonians (\ref{h1}), (\ref{h2}) and (\ref{h3}),
we find that our system
corresponds to the system (\ref{hg0}) with $k=1$
since the term with
function $M(\tilde W^2(x))$ can be removed by the
appropriate canonical transformation.
Therefore, we conclude
that the system (\ref{CalogeroH}) is canonically
equivalent to the $k=1$ supersymmetric system
(\ref{hg0}).

Analogously,  applying
the canonical transformation (\ref{CT})
to the system
(\ref{CalogeroH}) with $c>0$
($c=\gamma^2$, $\gamma\ne 0$), one
can reduce it  to the $k=0$
supersymmetric system  with the Hamiltonian of the
form
(\ref{decouple}).
Indeed, let us
define the new superpotential $\tilde W(x)$ by the
relation
$$
 \tilde W^2(x)+2\tilde v=W^2(x)
 -\frac{\gamma^2}{W^2(x)}+2v=2V(x),
$$
where without loss of generality we assume that
$W(x)>0$, and the constant $\tilde v$ has to
be chosen to provide the inequality
$\tilde W^2(x)\ge 0$. As a result, one can
represent  $W(x)$ in terms of the new superpotential as
$$
 W(x)=2^{-\frac 12}\sqrt{{\cal D}+\sqrt{{\cal D}^2
 +4\gamma^2}},\qquad {\cal D}=\tilde W^2(x)-2\Delta v,
$$
where $\Delta v=v-\tilde v$. Expressing $L(x)$ in terms
of $\tilde W$, we obtain
$L(x)\propto\tilde W'(x)\tilde W(x)M(\tilde W^2(x))$,
with the function $M(\tilde W^2(x))=[1-[{\cal D}^2+4
\gamma^2]^{-1/2}]\cdot W^{-\frac 12}(\tilde W(x))$ to
be regular due to the inequality $W(x)>0$. Therefore
the nilpotent term can be removed from the Hamiltonian  by
the canonical transformation\footnote{See Appendix for
the details.} (\ref{CT}) since the generating function
$G(p,\tilde W)$ is regular in this case.

Let us turn now to the next, $k=3$ case given by
the system of equations
\begin{equation}
\label{k=3}
 \varphi _3'-\frac 23y\left( \varphi _{ 2}'-y \right)-
 \varphi _2 =0, \quad
 \varphi _3+\frac 13\varphi _2\left(\varphi _2'
 -y \right) =0.
\end{equation}
Substituting $\varphi _3$ from the second equation into
the first one, we arrive at the nonlinear
differential equation of the second order
\[
 \varphi _2\varphi _2''+\left(\varphi _2'\right) ^2+y
 \varphi _2'+2\varphi _2-2y^2=0,
\]
which can equivalently be represented as the first
order
nonlinear differential equation
\[
 y\varphi _2\varphi _2'+\frac{1}{2}\left( y^2-\varphi _
 2\right) \left( y^2+\varphi _2\right) +y^2\left(
 \varphi _2-y^2\right) -C=0.
\]
When the integration constant $C=0$, the solution to
the
equation is a root of
the 4th order algebraic equation for $\varphi_2(y^2)$:
\[
 \left(3\varphi _2-y^2\right)\left(y^2+\varphi _2
 \right) ^3-C_1y^2=0.
\]
At $C_1=0$, its solutions have a simple form:
$\varphi _2=y^2/3 $, $\varphi _2=-y^2$. The first
solution corresponds to the holomorphic case with
$k=3$, while the second one  does to the case of $k=1$
supersymmetry with the supercharge
multiplied by the Hamiltonian. Even in the case $C=0$,
$C_1\neq 0$, the corresponding solutions
$\varphi _2(y^2)$ have a complicated form of solutions
in radicals of the 4th order equation, whereas for
$C\ne 0$ we have not succeeded in finding any
analytical solution of the nontrivial nature.
The same complications appear with
finding the nontrivial solutions to the system
(\ref{phn}) for $k>3$.

\section{Quantum anomaly and $k$-supersymmetry}
\label{quantum}

As we have seen in the previous section,  the
supersymmetry in classical one-dimensional system is
defined by the arbitrary function $W(x)$
and in general case the supercharges together with the
Hamiltonian form a nonlinear superalgebra.
According to the results of Ref. \cite{susy-pb} on the
supersymmetry in pure parabosonic systems,
a priori one can not exclude the situation
characterized by  the supercharges to be the
nonlocal operators represented in the form of some
infinite series in the operator $\frac{d}{dx}$.
Since such nonlocal supercharges
have to  anticommute  for some function of the Hamiltonian
being a usual local differential operator of the second order,
they have to possess a very peculiar
structure\footnote{In pure parabosonic systems
revealing $k$-supersymmetry both the supercharges
and Hamiltonian have a nonlocal structure
\cite{susy-pb}.}.
Due to this reason, we restrict ourselves by the discussion
of  the supersymmetric  systems with the
supercharges being the
differential operators of order $k$.
Classically
this corresponds to the system (\ref{hg0}) with the
holomorphic supercharges or to  the systems
discussed in Section \ref{c2}.

In Ref. \cite{susy-pb} it was observed that just in the
simplest
case of the superoscillator possessing the  nonlinear
$k$-supersymmetry and characterized by the holomorphic
supercharges of the form (\ref{hg0})
with the simplest superpotential $W(x)=x$,
the form of the classical superalgebra
$\{Q^+_k,Q^-_k\}=H^k$ is changed for
$\{Q^+_k,Q^-_k\}=H(H-\hbar)(H-2\hbar)\ldots
(H-\hbar(k-1))$ due to the quantum noncommutativity.
Moreover, it was also observed that for $W(x)\neq ax+b$
in generic case we have a global quantum anomaly
\cite{qa1,qa2}:
the direct quantum analogue of the superoscillators
loose the property of the conservation,
$[Q^\pm_k,H_k]\neq 0$.
Therefore, we arrive at the problem of looking for the
classes of superpotentials and corresponding quantization
prescriptions leading to  the quantum $k$-supersymmetric
systems without quantum anomaly.
This and  next two sections are devoted to the solution of
such a problem.

\subsection{The $k$-supersymmetry:
quadratic superpotential.}
\label{q1}

In this and next subsections we consider the quantization of
the nonlinear supersymmetry characterized by the
holomorphic form of the supercharges (\ref{hg0}).
As we shall see,  the straightforward
quantization without special quantum corrections
leads to the rigid restrictions on the form of the
superpotential $W(x)$.

Let us fix the quantum supercharges in the holomorphic
form corresponding to the classical $k$-supersymmetry,
\begin{equation}
\label{Qf}
 Q^\pm=2^{-\frac k2}(A^\mp)^k\theta^\pm,
\end{equation}
with
\begin{equation}\label{A}
  A^\pm={}\mp\hbar \frac{d}{dx}+W(x).
\end{equation}
With realization
\begin{equation}
\label{qn}
N=\theta^+\theta^-=
\frac{\hbar}{2}(\sigma _3+1),
\end{equation}
the operator $N$ has a sense of
the fermionic quantum
number and being normalized for $\hbar$
is the  projector onto the fermionic subspace,
whereas the complimentary projector onto the bosonic
subspace is $\theta^-\theta^+/\hbar=1-N\hbar^{-1}$.
Then choosing  the quantum Hamiltonian in the form
(\ref{Hgen}), from the
requirement of conservation of the supercharges,
$\left[Q^\pm,\,H\right]=0$, we arrive at  the equations
\begin{eqnarray}
 &L(x)=kW'(x),&\label{lkw}\\
 &V(x)=\frac 12\left(W^2(x)-k\hbar W'(x)\right)+v,&
 \label{quadro} \\
 &\hbar^3 k(k^2-1) W'''(x)=0.&
 \label{hw}
\end{eqnarray}
Therefore,  the quantum system given by the Hamiltonian
\begin{equation}
\label{Hq1}
H=\frac{1}{2}\left(-\hbar^2\frac{d^2}{dx^2}+W^2(x)+
2v+k\hbar\sigma _3W'\right)
\end{equation}
possesses the nonlinear supersymmetry of order $k\geq 2$
characterized by the  holomorphic supercharges (\ref{Qf})
only when
\begin{equation}
\label{w2}
W(x)=w_2x^2+w_1x +w_0.
\end{equation}
For any other form of the superpotential the nilpotent
operators (\ref{Qf}) are not conserved that can be treated
as a quantum anomaly. Below we shall see that the
quantum anomaly for the system (\ref{Hq1}) can be
``cured" for some superpotentials  if to  modify
appropriately (by  $\hbar$-dependent terms) the
supercharges.
It is necessary to stress that the relation (\ref{hw})
fixing the form of the superpotential for $k\geq 2$
has a purely quantum nature.
One also notes that with the prescription
(\ref{qn}),  $V(x)$ given by Eq. (\ref{quadro})  plays the
role of  the total potential  for the bosonic sector,
$V(x)=V_B(x)$, whereas the potential
for the fermionic sector is $V_F(x)=V(x)+\hbar kW'(x)$.
The appearance in $V(x)$ of the term proportional to
$\hbar$ can obviously be associated with the ordering
ambiguity under construction of the operator
$N$: under the choice $N=\frac{1}{2}[\theta^+,\theta^-]$
instead of (\ref{qn}), we again arrive at the Hamiltonian
(\ref{Hq1}), but the term linear in $\hbar$ dissappears
from $V(x)$. In what follows we
shall have in mind the quantum prescription (\ref{qn}).

The anticommutator of supercharges $Q^+$ and $Q^-$
gives the polynomial of order $k$ in $H$, e.g.,
for the simplest cases $k=2$,
$k=3$ and $k=4$ we have
\begin{eqnarray}
 &\left\{Q^-,Q^+\right\}=
 H^2-\frac 14\left(w_1^2-4w_0w_2\right) ,&\nonumber\\
& \left\{Q^-,Q^+\right\}=H^3-\left(w_1^2-4w_0w_2
 \right)H+2w_2^2,&\nonumber\\
& \left\{Q^-,Q^+\right\}=H^4-\frac 52\left(w_1^2
 -4w_0w_2\right)H^2+12w_2^2H+\frac{9}{16}w_2^2
 \left(w_1^2-4w_0w_2\right)^2,&\nonumber
 \end{eqnarray}
where for the sake of simplicity we have put $v=0$ and
$\hbar=1$.
The family of supersymmetric systems (\ref{w2})  is
reduced to  the superoscillator at $w_2=0$
with the associated exact $k$-supersymmetry
\cite{susy-pb}.
For $w_2\neq 0$ the $k$-supersymmetry is realized
always in the spontaneously broken phase
since in this case  the  supercharges (\ref{Qf}) have no zero
modes (normalized eigenfunctions of zero eigenvalue).

It is worth  noting that the nonlinear supersymmetry
with quadratic superpotential (3.8) was found
earlier in ref. \cite{AoyamaNP} as a
by-product in the context of the  discussion of the
non-renormalization theorem for supersymmetric theories.

\subsection{The $k$-supersymmetry:  exponential
superpotential}
\label{q2}

Let us take  the supercharges in the form of  polynomials of
order $k$ in the oscillator-like variables $A^\pm$
(\ref{A}):
\begin{equation}\label{QP}
 Q^\pm=2^{-\frac k2}\left\{(A^\mp)^k+\sum_{n=0}^{k-
 1}
 q_{k-n}(A^\mp)^n\right\}\theta^\pm,
\end{equation}
where $q_n$ are real parameters  have to be fixed.
If we treat the supercharges (\ref{QP}) classically,
then the condition of their conservation by the Hamiltonian
of the general form (\ref{Hgen}) results in
$$
Q^\pm=2^{\frac{k}{2}}\left(A^\mp +
\frac{q_1}{k}\right)^k\theta^\pm,\quad
H=\left(\{Q^+,Q^-\}\right)^{\frac{1}{k}}+v,
$$
i.e. we arrive again at  the $k$-supersymmetric system
(\ref{hg0}) with the arbitrary function $W(x)+q_1/k$
playing the role of the superpotential.
However, as we shall see, quantum mechanically
ansatz  (\ref{QP}) gives us a nontrivial  family of
$k$-supersymmetric systems related to the so called
quasi-exactly solvable problems
\cite{turbiner,shifman,ushv,olv,dunne}.
First one notes that  the parameter $q_1$ can be
removed by a simple shift  of the superpotential
both on the classical and quantum levels and  we
can put $q_1=0$.
Then, as in the case of the supercharges (\ref{Qf}),
the requirement of conservation of (\ref{QP})
results in  the Hamiltonian (\ref{Hq1})
as well as in  some $k-2$ algebraic equations  fixing
the parameters $q_3$, $q_4$, $\ldots$, $q_k$
in terms of $q\equiv q_2$,
whereas  the condition (\ref{hw}) for $k\geq 2$
is changed now for
\begin{equation}
\label{eWExp}
 \hbar^3W'''-\omega_k^2\hbar W'=0,\qquad
 \omega_k^2=-\frac{24q}{k\left(k^2-1\right)}.
\end{equation}
This means that the supercharges (\ref{QP})
contain only  the one-parameter arbitrariness and
the case of quadratic superpotential (\ref{w2})
is included here as a particular case corresponding to
$q=0$.
For  $q\ne 0$ the solution
of Eq.  (\ref{eWExp}) acquires the exponential form
\begin{equation}
\label{WExp}
 W(x)=w_+e^{\omega _kx}+w_-e^{-\omega _kx}+w_0,
\end{equation}
where all the parameters $w_{\pm,0}$ are real, while
the parameter $\omega_k$ is real or pure imaginary
depending on the sign of $q$,
and for the sake of simplicity we put $\hbar=1$.
In the limit
$\omega_k\to 0$ this superpotential is
reduced to the quadratic form (\ref{w2})
via the  appropriate rescaling of the parameters
$w_{\pm,0}$ .

For $k=2$ the anticommutator of the supercharges has
the form
\[
 \left\{Q^-,\,Q^+\right\} =\left(H-c_-\right)
 \left(H-c_+ \right),
\]
where
\[
 c_\pm ={}\pm\omega _2\sqrt{w_0^2-4w_+w_-}+q+v.
\]
For  $k\geq 2$ the superalgebra for the system with
superpotential (\ref{WExp}) can be represented as
\begin{equation}
 \left\{ Q^-,\,Q^+\right\} =H^k+\sum _{n=1}^kc
 _nH^{k-n}, \qquad c_n \in \mathbb R.  \label{P(H)}
\end{equation}
In principle, the coefficients $c_n$ can be found explicitly
since the system of linear equations arises for them.
The values of energy of the supercharges' singlets
are the roots of the polynomial on the right hand side
of Eq. (\ref{P(H)}) and to
find them one has to solve the algebraic equation of
the corresponding order.
In next section we analyse in detail the class of $k$-
supersymmetric systems given by the superpotential
(\ref{WExp}) in the context of the partial algebraization
of the spectral problem.


\section{The $k$-supersymmetry  and partial algebraization
of the spectral problem}
\label{algebraiz}

In the late eighties  a new class of spectral problems
was discovered \cite{turbiner,shifman}.
It occupies the  intermediate position between the
exactly solvable problems and all others.
According to Ref. \cite{shifman1},
the quantum-mechanical system is quasi-exactly solvable or
admits partial algebraization of the spectrum
if its potential depends explicitly on the natural
parameter  $n$ in such a way that
exactly $n$ levels in the spectrum  can be found
algebraically.
The unique nature of such systems is
based on the hidden symmetry of the Hamiltonian.
The part of the spectrum that can be found algebraically is
related to finite-dimensional representations of
the corresponding Lee group (algebra).
For one-dimensional systems  the
quasi-exact solvability is associated with
non-unitary finite-dimensional representations of the
$sl(2,\mathbb R)$. Such representations are
characterized by a parameter $j$ referred to as a ``spin'',
which can take integer and half-integer values. The
number of the eigenstates of the Hamiltonian that can be
found algebraically is equal to $2j+1$.

Here we argue in favour of existence of the  intimate
relation between nonlinear supersymmetry and the partial
algebraization
scheme \cite{shifman, shifman1}. For example, if in a
given system with the nonlinear supersymmetry of the
order $k$,
\[
 \{Q^-,Q^+\}=\left(H-E_1\right)\ldots\left(H-E_k\right),
\]
there are $k$ singlets in the bosonic or fermionic
sectors, i.e. $k$ zero modes of $Q^+$ or $Q^- $, then
the eigenvalues of the corresponding  states are equal to
$E_i$. Then it is quite obvious that if such a  system
is not exactly-solvable, it admits the
partial algebraization of its spectrum.
Having in mind these preliminary comments, let us
show that the supersymmetric system with
the superpotential (\ref{WExp})
can be related to some known families of
quasi-exactly solvable problems.

Let us put $\hbar =1$ and consider the potentials
\begin{eqnarray}
 V(x)&=&\frac 12\left(ae^x+b-2j\right)^2+\frac 12
 \left(ce^{-x}+b+1\right)^2, \label{e^x} \\
 V(x)&=&\frac{a^2}2\cos^2x+a\left(2j+\tfrac 12\right)
 \sin x, \label{cos x} \\
 V(x)&=&\frac{a^2}2\sinh^2x-a\left(2j+\tfrac 12\right)
 \cosh x, \label{cosh x} \\
 V(x)&=&\frac{a^2}2\cosh^4x-\frac a2\left(a+4j+2\right)
 \cosh^2x, \label{cosh4}
\end{eqnarray}
admitting  the
partial algebraization of the spectrum \cite{shifman,
shifman1}.
The potential (\ref{quadro}) with the
superpotential of the from (\ref{WExp}) coincides
with the potential (\ref{e^x}) when
\begin{gather*}
 \omega _k=1,\quad w_+=a,\quad w_-=c,\quad w_0=1+2
 \left(b-j\right) ,\quad k=2j+1, \\
 v=\frac 34+b+b^2-2ac+j-2bj+3j^2,
\end{gather*}
or with the potential (\ref{cos x}) when
\begin{equation}\label{oddk1}
 \omega _k=-i,\quad w_-=w_+=\frac a2,\quad
 w_0=0, \quad v= 0,\quad k=4j+1,
\end{equation}
or with the potential (\ref{cosh x}) when
\begin{equation}\label{oddk2}
 \omega _k=1,\quad w_\pm=\pm\frac a2,\quad
 w_0=0,\quad v=0,\quad k=4j+1,
\end{equation}
or with the potential (\ref{cosh x}) when
\[
 \omega _k=2,\quad w_\pm=\pm\frac a4,\quad
 w_0=0,\quad v={}-a\left(j+\tfrac 12\right),
 \quad k=2j+1.
\]

In the cases
(\ref{oddk1}) and (\ref{oddk2}), there is a complete
correspondence with the quasi-exactly solvable
potentials (\ref{cos x}) and (\ref{cosh x}) for odd
$k$ only. Having in mind this observation, we  first analyse
in detail  the correspondence between the nonlinear
supersymmetry and the
partial algebraization scheme for  the case of quasi-exactly
solvable
potential (\ref{cosh x}).
Writing down this potential as
\begin{equation}\label{cosh}
 V(x)=\frac{a^2}2\sinh^2x-\frac{a}2k\cosh x,
 \qquad k\in\mathbb N,
\end{equation}
we see that it has exactly the form of the potential
(\ref{quadro}) of the nonlinear supersymmetry
system with the
corresponding superpotential
\begin{equation}
 W(x)=a\sinh x.
 \label{Wsh}
\end{equation}
Since  the potential of the form
(\ref{cosh}) corresponds to the potential
(\ref{cosh x}) for odd $k$ only, let us investigate the
$k$-supersymmetric systems with even $k$
and start from the case $k=2$. Formally this
corresponds to the ``spin" $j=1/4$ in the partial
algebraization scheme \cite{shifman} for the
potential (\ref{cosh x}). In this case $q=-1/4$ and
the supercharges read as
\[
 Q^\pm=\frac 12\left((A^\mp)^2-\frac 14\right)\theta^\pm,
\]
where $A^\pm$ are defined in (\ref{A}). Zero modes
of the supercharge $Q^+$ belong to the bosonic sector
and have the form
\[
 \varphi_\pm(x)=e^{\pm\frac x2}e^{-a\cosh x},
\]
where for definiteness we have supposed that $a>0$.
The anticommutator of the supercharges is
\[
 \left\{Q^-,Q^+\right\}=\left(H-\frac a2+\frac 18
 \right)\left(H+\frac a2+\frac 18\right),
\]
and we find that the  linear combinations
$
 \varphi^{(\pm)}=\varphi _-\pm\varphi _+
$
are the eigenfunctions  of the bosonic Hamiltonian,
\begin{equation}
 H_B\varphi=E\varphi ,\label{H_B}
\end{equation}
corresponding to the eigenvalues
\[
E_\pm ={}\mp \frac a2-\frac 18.
\]

The fermionic Hamiltonian $H_F$ is characterized by  the
potential
\[
 V_F(x)=\frac{a^2}2\sinh^2x+\frac a2 k\cosh x,
\]
which is the superpartner of the potential
(\ref{cosh}). Since this differs from
(\ref{cosh}) in the sign before the second term,
the corresponding solutions in the fermionic sector
in the case of $a<0$ can be obtained from the bosonic
solutions with $a>0$ by a simple change $a\rightarrow -a$.
Therefore,  we see that there are two bound states in the
bosonic (fermionic) sector for $a>0$ ($a<0$),
that means that
the potential (\ref{cosh}) is quasi-exactly solvable
for $k=2$ as well. As we shall  see,  the same conclusion
is also true  for any even $k$.

Let us consider the case $k=3$ corresponding to the
spin $j=1/2$. In this case the supercharges $Q^\pm$
acquire the form
\[
 Q^\pm=2^{-\frac 32}A^\mp\left(\left(A^\mp\right)^2
 -1\right)\theta ^\pm.
\]
The zero modes of the operator $Q^+$ are
\[
 \varphi _1=e^{-a\cosh x},\qquad
 \varphi _{2,3}=e^{\pm x}e^{-a\cosh x}.
\]
Looking for the  solutions to the equation (\ref{H_B}) in
the form of the linear combination of the zero modes,
$
 \varphi =c_1\varphi _1+c_2\varphi _2+c_3\varphi _3,
$
we obtain the following three (not normalized)
eigenfunctions
$$
 \varphi_0=\sinh x\,e^{-a\cosh x},\qquad
 \varphi_\pm=\left(1+c_\pm\cosh x\right)
 e^{-a\cosh x},
 $$
 where
 $c_\pm=(4a)^{-1}\cdot (1\mp\sqrt{16a^2+1}) $.
 The energies of these states are
\[
 E_0=-\frac 12,\qquad E_\pm=\frac 14\left({}
 \pm\sqrt{16a^2+1}-1\right) ,
\]
and the anticommutator of the supercharges is
\begin{equation}\label{aQ3}
 \left\{Q^-,\,Q^+\right\}=\left(H-E_0\right)
 \left(H-E_-\right)\left(H-E_+\right).
\end{equation}
It is necessary to stress that though in the case $k=3$
the direct application of
the partial  algebraization scheme to the potential allows
ones to find only two eigenstates and corresponding
eigenvalues
($\varphi _\pm$, $E_\pm$),
the concept of nonlinear supersymmetry  gives the
information on
one more  exact eigenstate ($\varphi _0$) of the
Hamiltonian.
The same  is also valid for any odd $k$: the algebraization
scheme gives for the potential (\ref{cosh}) the information
on $\frac{k+1}{2}$ eigenvalues  corresponding to the
even
(in $x$) eigenstates, whereas
the nonlinear supersymmetry allows ones  to find
in addition  $\frac{k-1}{2}$ eigenvalues  corresponding
to the odd eigenfunctions.

The  eigenfunctions and eigenvalues in  the
fermionic sector for $a<0$ can be obtained via  the
formal change  $a\to-a$, and we conclude that
in the case $k=3$ there are  three bound states
in bosonic (fermionic) sector for $a>0$ ($a<0$).

Now let us generalize  the $k$-supersymmetry  of
the system given by the superpotential (\ref{Wsh})
for the case of arbitrary $k$.
Using  the explicit  form of the supercharges for
$k=2,\, 3 $,  one can
suppose that the supercharges $Q_k^\pm$ obey the
recurrent relation
\begin{equation}
\label{Qrec}
 Q_{k+2}^\pm=\frac 12\left(\left(A^\mp\right)^2
 -\left(\tfrac{k+1}2\right)^2\right)Q_k^\pm.
\end{equation}
Let us prove the conservation of the supercharges
(\ref{Qrec})
by the method of mathematical induction using this
conjecture. As the first step we suppose that for some
given $k$ the supercharge is conserved. Thus,
calculating $[Q_k^+,\,H_k]$, we obtain  the relation
\begin{equation}
 \left[W',Q_k^+\right]+kWA^-Q_k^++\frac 14
 \left(k-1\right)kW'Q_{k-2}^+=0.  \label{BaseRec}
\end{equation}
The Hamiltonian $H_{k+2}$ is related to the $H_k$ as
$
 H_{k+2}=H_k-W'+2W'N.
$
Then we have
\[
 \left[Q_{k+2}^+,\,H_{k+2}\right]=\left[W',\,Q_{k+2}^+
 \right]+\left(k+2\right)WA^-Q_{k+2}^++\frac 14
 \left(k+1\right)\left(k+2\right)W'Q_k^+.
\]
Using the relations (\ref{Qrec}) and (\ref{BaseRec}),
after rather cumbersome algebraic manipulations one can
reveal that the expression vanishes. Since the corresponding
supercharges are conserved for $k=2,\, 3$,
we conclude that  the supercharges $Q_k^\pm$ are
conserved for any $k$ as well. This also proves the
conjecture (\ref{Qrec}).

The relation (\ref{Qrec}) gives us the following
form of the supercharges for arbitrary  $k$:
\[
 Q_k^\pm=2^{-\frac k2}
 \left(A^\mp+\tfrac{k-1}2\right)
 \left(A^\mp+\tfrac{k-3}2\right)\ldots
 \left(A^\mp-\tfrac{k-3}2\right)
 \left(A^\mp-\tfrac{k-1}2\right)\theta^\pm.
\]
Using this representation, we find all the zero
modes of the supercharge $Q_k^+$:
\[
 \varphi_s(x)=e^{sx-a\cosh x},\qquad s=
 {}-\tfrac{k-1}2,\,-\tfrac{k-3}2,\,\ldots,\,
 \tfrac{k-3}2,\,\tfrac{k-1}2.
\]
Since these modes belong to the bosonic sector, we can
build $k$ eigenfunctions of the bosonic part of the
Hamiltonian as a linear combination of them:
\[
 \psi(x)=e^{-a\cosh x}\left\{
 \begin{array}{ll}
  \displaystyle\sum_{n=-m}^mc_ne^{nx},&\quad
  k=2m+1,
  \\[4mm]
  \displaystyle\sum_{n=-m}^{m-1}c_ne^{\left(
  n+\frac 12\right)x},&\quad k=2m.
 \end{array}
 \right.
\]
If we put  this  into the
corresponding stationary Schr\"odinger equation,
we arrive at   the recurrent system of algebraic equations
on energy $E$ and the coefficients $c_n$
\[
 \frac 1a\left(n^2+2E\right)c_n+c_{n+1}\left(m+n+1
 \right)+c_{n-1}\left(m-n+1\right)=0,\qquad\text{for
 odd }k,
\]
where $c_n=0$ for $\left| n\right|>m$, and
\[
 \frac 1a\left( \left( n+\tfrac{1}{2}\right) ^2+2E
 \right) c_n+c_{n+1}\left( m+n+1\right) +c_{n-1}\left(
 m-n\right) =0, \qquad \text{ for even }k,
\]
where $c_n=0$ for $n<-m$ or $n\geq m$.
Thus, we have demonstrated how the concept of nonlinear
supersymmetry allows us to reduce the problem of finding
the part of the spectrum for the potential (\ref{cosh})
to the pure algebraic problem.

Let us discuss shortly  the potential (\ref{cosh4}), to which
the  superpotential $W(x)=\frac a2\sinh 2x$
corresponds within the framework of the
nonlinear supersymmetry. {}From the form of the
superpotential one can
immediately conclude that the potential
(\ref{cosh4}) can be reduced to the
form (\ref{cosh}) just by rescaling the argument and
the parameter $a$. Indeed, with the help of the
relation
\[
 \frac{a^2}2\cosh^4x-\frac a2\left(a+4j+2\right)
 \cosh^2x=\frac{a^2}8\sinh^22x-\frac{2j+1}2a
 \cosh 2x-\frac{2j+1}2a,
\]
after rescaling $x\to 2x$, $a\to a/4$, we arrive at
the form (\ref{cosh x}). But in the  case of the potential
(\ref{cosh4}), there is  a
difference in comparison with (\ref{cosh x}).
As we have mentioned  above, according to
Ref.  \cite{shifman, shifman1},  only
the even eigenstates of Hamiltonian with the potential
(\ref{cosh x}) can be found following the
$sl(2,\mathbb R)$ partial algebraization scheme, whereas
for the potential (\ref{cosh4}) all the lowest $2j+1$
states can be found  within the framework of the very
scheme.
But we see that though the potentials (\ref{cosh x})
and (\ref{cosh4}) have different representations in the
algebraization scheme, in reality  they correspond to
the same physical system. Moreover,  we see that all the
potentials (\ref{e^x})-(\ref{cosh4}) are particular  cases
of the potential (\ref{quadro}) with the exponential
superpotential (\ref{WExp}). For the potential
(\ref{quadro}) the first
$k$ states can be found algebraically with the help of the
associated  nonlinear supersymmetry.
In other words, the nonlinear supersymmetry provides us
with a universal point of view on the quasi-exactly
solvable potentials (\ref{e^x})-(\ref{cosh4}).

Now let us discuss the superpotential (\ref{WExp}) of
the general form, which, as we have seen, allows ones to
unify  all the cases (\ref{e^x})-(\ref{cosh4}). The potential
(\ref{quadro}) with this superpotential has the form
\[
 V(x)=\frac 12w_+^2e^{2\omega x}+\left(w_0-\tfrac k2
 \omega\right)w_+e^{\omega x}+\left(w_0+\tfrac k2
 \omega\right)w_-e^{-\omega x}+\frac 12w_-^2
 e^{-2\omega x}+v+w_+w _-+\frac 12w_0.
\]
The superpartner of this potential can be obtained
by the formal substitution $k\to-k$.
Consider in detail the simplest case $k=2$. The
corresponding supercharge
\[
 Q^+=\frac 12\left((A^-)^2-\frac{\omega^2}4\right)
 \theta ^+
\]
has the zero modes
\begin{equation}
 \varphi _\pm=\exp\left(\left({}\pm\tfrac\omega 2
 -w_0\right)x-\tfrac 1\omega\left(w_+e^{\omega x}
 -w_-e^{-\omega x}\right)\right) ,  \label{ph}
\end{equation}
and   for the equation
$H_B\varphi=E\varphi$ one finds the two eigenfunctions
\[
 \varphi^{(\pm)}=\varphi_-+c_\pm\varphi_+,\qquad
 c_\pm=\frac 1{2w_-}\left(w_0\pm
 \sqrt{w_0^2-4w_+w_-}\right),
\]
with the energy
\[
 E_\pm={}\pm\frac 12\omega
 \sqrt{w_0^2-4w_+w_-}+v-\frac{\omega^2}8.
\]
In the fermionic sector, the zero modes of the
supercharge $Q^-$ are
\begin{equation}\label{psi}
 \psi_\pm=\exp\left(\left({}\pm\tfrac\omega 2
 +w_0\right)x+\tfrac 1\omega\left(w_+e^{\omega x}
 -w_-e^{-\omega x}\right)\right)
\end{equation}
and all the corresponding formulas can by reproduced
via the changes $w_\pm\to-w_\pm$, $w_0\to-w_0$.
As it follows from (\ref{ph}) and (\ref{psi}),  in the case
$w_+w_-<0$ there are two bound states in bosonic
(fermionic) sector when $w_+>0$ ($w_+<0$),
whereas for $w_+w_->0$ there are no such bound states
at all.
The corresponding $k=2$ superalgebra is written as
\[
 \left\{ Q^-,Q^+\right\}=\left(H-\tfrac 12q\right)^2
 +q\left(w_0^2-4w_+w_-\right).
\]

Using the obtained results for the potential (\ref{cosh})
and realizing the substitution  $x\to\omega x$,
we arrive at the supercharges
obeying the recurrent relation
\[
 Q_{k+2}^\pm=\frac 12\left(\left(A^\mp\right)^2-
 \left(\tfrac{k+1}2\right)^2\omega^2\right)Q_k^\pm.
\]
{}From the consideration of the nonlinear supersymmetry
with the
superpotential (\ref{WExp}), it  follows that for any
$k$ the supercharges depend on the parameter $\omega_k
$
and do not depend explicitly on other parameters of the
superpotential. This means that the supercharges are
conserved and obey the same recurrent relation if we
rescale the parameter $q$ to yield the relation
$\omega_k=\omega$.
Then in the same way as for the potential (\ref{cosh}),
one can look for the wave functions of the $k$ singlets
in the form
\[
 \psi(x)=e^{-\int^xW(y)\,dy}\left\{
 \begin{array}{ll}
  \displaystyle\sum_{n=-m}^mc_ne^{n\omega x},
  &\quad k=2m+1, \\[4mm]
  \displaystyle\sum_{n=-m}^{m-1}c_ne^{\left(
  n+\frac 12\right)\omega x}, &\quad k=2m.
 \end{array}
 \right.
\]
The corresponding stationary Schr\"odinger equation
gives us the recurrent system of algebraic equations on
the energy and on the coefficients $c_n$:
\[
 c_n(E-v+\tfrac 12n\omega(n\omega-2w_0))+(m-n+1)
 \omega w_+c_{n-1}-(m+n+1)\omega w_-c_{1+n}=0,
\]
for odd $k$, where $c_n=0$ for $|n|>m$, and
\[
 c_n(E-v+\tfrac 12(\tfrac 12+n)\omega((\tfrac 12+n)
 \omega-2w_0))+(m-n)\omega w_+c_{n-1}-(m+n+1)
 \omega
 w_-c_{n+1}=0,
\]
for even $k$, where $c_n=0$ for $n<-m$ or $n\geq m$.
Excluding the coefficients $c_n$ from the recurrent
systems, one can obtain the algebraic equation for $E$
of the order $k$. Normalizing the corresponding
polynomial to the form $E^k+\ldots$ and substituting
$E\to H$, one can get the exact form of the polynomial
specifying the $k$-supersymmetry via
the anticommutator of the supercharges
$Q^\pm$.
Similarly to the case $k=2$,
one can find that if  $w_+w_-<0$, then
there are $k$ bound states in bosonic (fermionic)
sector when $w_+>0$ ($w_+<0$). If $w_+w_->0$, such
bound singlets do not exist at all.

To conclude this  section, we note that
the choice of the parameters in the
superpotential (\ref{WExp})
\begin{gather*}
 \omega _k=\alpha,\quad w_+=0,\quad w_-=\pm B,\quad
 w_0={}\mp A-\frac\alpha 2\left(k\pm 1\right), \\
 v=-\frac 18\alpha\left(k\pm 1\right)\left(
 \left(k\pm 1\right)\alpha\pm 4A\right)
\end{gather*}
leads to the exactly solvable Morse potential
\cite{cooper}
\begin{equation}\label{CalogeroP}
 V(x)=A^2+B^2e^{-2\alpha x}-B(2A+\alpha )e^{-\alpha
 x},
\end{equation}
where $A$, $B$ are real positive parameters. As it is well
known, from the Morse potential a
vast number of famous exactly solvable potentials can
be reproduced by the point canonical transformations and
suitable limiting procedures \cite{cooper}. This
illustrates an intimate though indirect relation of
the nonlinear supersymmetry  not only to the quasi-exactly
solvable systems but to the exactly solvable models as
well.

\section{Anomaly-free  $k=2$ supersymmetry}
\label{anomaly}

In the subsection \ref{PolynomCase},
we have analysed the classical supersymmetry from the
viewpoint of the polynomial in momentum structure
of the supercharges. Since for the second order
supercharges the solution was found by us in the general
form, let us  analyse  this case at the quantum level too.
On the other hand, there is no sense to analyse in such a
context the orders higher then 2  since we
do not know the general solution even at classical
level. Different aspects of supersymmetry with the
general second order supercharges were considered in
Ref.~\cite{andrian}. Here we discuss the general
quantum case of the $k=2$ supersymmetry  in
the context of ``curing" the quantum anomaly problem
and from the viewpoint of all the possible types  of spectra.
Then, as an example of application  of
the $k=2$ supersymmetry, we
construct a   new exactly solvable
nontrivial quantum model associated with the
infinitely deep well.

\subsection{Quantum $k=2$ supersymmetry vs.
classical supersymmetry}
\label{vs}

Let us consider the second order supercharge of the
general form
\begin{equation}
 Q^+=\frac 12\left(\hbar^2\frac{d^2}{dx^2}+2\hbar f(x)
 \frac{d}{dx}+b(x)\right)
 \theta^+, \label{Qd2}
\end{equation}
and the most general quantum Hamiltonian
\[
 H={}-\frac{\hbar^2}{2}\frac{d^2}{dx^2}
 +V(x)+L(x)\theta^+\theta^-.
\]
The condition $[Q^+,\,H]=0$ partially determines the
unknown functions in $Q$ and $H$:
\begin{eqnarray}
 L(x)&=&2f'(x), \label{l2q}\\
 V(x)&=&\frac 12\left(f-\frac{\hbar f'}{2f}\right)^2
 -\frac{\hbar} 2\left(f-\frac{\hbar f'}{2f}\right)'-
 \frac{c}{2f^2}+v, \label{Vk=2} \\
 \label{b2q}
 b(x)&=&\frac c{f^2}+f^2+\hbar f'-\frac{\hbar^2f''}{2f}
 +\left(\frac{\hbar f'}{2f}\right)^2,
\end{eqnarray}
where $c$ and $v$ are real constants.
The function  (\ref{Vk=2}) plays the role of
the potential for the bosonic sector.
Taking into account Eqs. (\ref{l2q})--(\ref{b2q})
and identifying $f(x)=W(x)$,
we obtain the following most general form for the
Hamiltonian and the supercharge of the quantum
$k=2$ supersymmetry:
\begin{eqnarray}
H&=&\frac{1}{2}\left[-\hbar^2\frac{d^2}{dx^2}+
W^2-\frac{c}{W^2}+2v+
2\hbar W'\sigma_3+\Delta(W)\right] \label{HD}\\
Q^+&=&\frac{1}{2}
\left[\left(\hbar\frac{d}{dx}+W\right)^2+\frac{c}{W^2}
-\Delta(W)\right]\theta^+,\label{QD}\\
\Delta&=&\frac{\hbar^2}{4W^2}
\left(2W''W-W'{}^2\right).
\label{WD}
\end{eqnarray}
Looking at  the quantum Hamiltonian  (\ref{HD})
and supercharge (\ref{QD}),
we see that their form is different from the
direct quantum analogue constructed proceeding from the
classical quantities  (\ref{CalogeroH}) and
(\ref{CalogeroQ}) via the quantization prescription
$N=\frac{1}{2}[\theta^+,\theta^-]$:
the presence of quadratic in $\hbar^2$ term (\ref{WD})
in both operators $H$ and $Q^+$
is crucial for preserving the supersymmetry
at the quantum level. Therefore, one can say that
the quantum correction (\ref{WD}) cures the
problem of the quantum anomaly since without it the
supercharge would not be the integral of motion.
It is interesting to note that the quantum term of the form
(\ref{WD})
appears also in the method of constructing new solvable
potentials from the old ones via the operator
transformations \cite{cooper}.

The supercharges $Q^+$ and $Q^-=(Q^+)^\dagger$
satisfy the relation
\begin{equation}
 \left\{Q^+,\,Q^-\right\}=\left(H-v\right)^2+c
 \label{SAk=2}
\end{equation}
being the exact quantum analogue of the first
classical relation from Eq. (\ref{qqh}).
\begin{figure}
\begin{center}
\begin{picture}(280,100)
\put(-10,-10){\line(1,0){310}}
\put(-10,110){\line(1,0){310}}
\put(-10,-10){\line(0,1){120}}
\put(300,-10){\line(0,1){120}}

\put(0,10){\vector(1,0){50}}
\put(10,0){\vector(0,1){100}}
\put(25,-5){a)}
\put(20,30){\circle* 4}
\put(30,30){\circle* 4}

\put(20,50){\circle* 4}
\put(30,50){\circle* 4}

\put(20,70){\circle* 4}
\put(30,70){\circle* 4}

\put(20,90){\circle* 4}
\put(30,90){\circle* 4}

\put(105,-5){b)}
\put(80,10){\vector(1,0){50}}
\put(90,0){\vector(0,1){100}}
\put(100,10){\circle 4}

\put(100,30){\circle* 4}
\put(110,30){\circle* 4}

\put(100,50){\circle* 4}
\put(110,50){\circle* 4}

\put(100,70){\circle* 4}
\put(110,70){\circle* 4}

\put(100,90){\circle* 4}
\put(110,90){\circle* 4}

\put(185,-5){c)}
\put(160,10){\vector(1,0){50}}
\put(170,0){\vector(0,1){100}}
\put(180,10){\circle 4}
\put(190,10){\circle 4}

\put(180,30){\circle* 4}
\put(190,30){\circle* 4}

\put(180,50){\circle* 4}
\put(190,50){\circle* 4}

\put(180,70){\circle* 4}
\put(190,70){\circle* 4}

\put(180,90){\circle* 4}
\put(190,90){\circle* 4}

\put(265,-5){d)}
\put(240,10){\vector(1,0){50}}
\put(250,0){\vector(0,1){100}}
\put(260,10){\circle 4}
\put(260,30){\circle 4}

\put(260,50){\circle* 4}
\put(270,50){\circle* 4}

\put(260,70){\circle* 4}
\put(270,70){\circle* 4}

\put(260,90){\circle* 4}
\put(270,90){\circle* 4}
\end{picture}
\end{center}
\caption{The four types of the spectra for  the $k=2$
supersymmetry.}
\label{k2sp}
\end{figure}
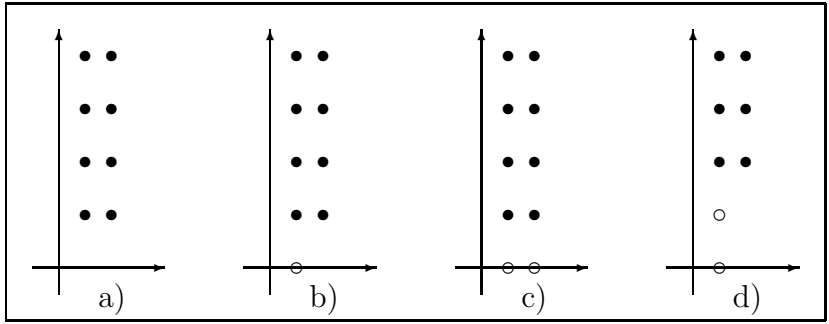
This superalgebra defines the form of the spectrum of
bounded states. For the sake of definiteness,
let us put the constant $v=0$.
Then in general the spectra of the $k=2$ supersymmetric
system can be  of the four types presented on
Fig.~\ref{k2sp}.
For  $c>0$, there
are no singlets in the system and we have completely
broken $k=2$ SUSY, see  Fig.~\ref{k2sp}${\bf a}$.
This case corresponds, in particular,
to the  $k=2$ SUSY given by  the quadratic
superpotential discussed in   Section \ref{q1}.

In the case $c=0$, there are three
possibilities: {\bf (i)} the completely broken phase (there
are
no  singlet states),
see Fig.~\ref{k2sp}${\bf a}$;
{\bf (ii)}
there is one singlet state in either bosonic or fermionic
sector, see Fig.~\ref{k2sp}${\bf b}$; {\bf (iii)} there are
two singlet states with equal energy,
one  in bosonic and another in fermionic
sector, see  Fig.~\ref{k2sp}${\bf c}$.
The  types of the spectra ${\bf b}$ and ${\bf c}$ are
represented by the new exactly solvable
model in  the next subsection.

In the case $c<0$,
all the three mentioned types of the spectra, ${\bf a}$,
${\bf b}$ and
${\bf c}$, can be realized and, in addition,
another situation with two
singlet states in one of the two (bosonic or fermionic)
sectors can exist, see
Fig.~\ref{k2sp}${\bf d}$.
Examples of the  type ${\bf d}$ were
represented above exhaustively by the
supersymmetric systems associated with the exponential
superpotential (\ref{WExp}).

\subsection{The $k=2$ supersymmetry in action:  a new
exactly solvable model}
\label{action}

Via the  appropriate choice of the superpotential,
one can construct  the $k=2$ supersymmetric systems
associated with the  exactly solvable potentials.
E.g., the simplest
choice of the superpotential $W(x)=x$ leads to the
famous (two-particle) Calogero model.
The Morse potential
(\ref{CalogeroP}) can also be reproduced and
there are two possibilities to realize it here:
\begin{align*}
 W(x)&=A-\frac 12\alpha-Be^{-\alpha x},&
 c&=-\frac 1{16}\alpha^2\left(2A-\alpha\right)^2,
 \\\intertext{or}
 W(x)&=Be^{-\alpha x}-A-\frac 32\alpha,&
 c&=-\frac 1{16}\alpha^2\left(2A+3\alpha\right)^2.
\end{align*}

It is well-known that the usual linear
supersymmetry allows ones  to obtain new exactly solvable
potentials from the  given ones. Here we show that the
$k=2$ supersymmetry can be exploited in the same
way. For example,  one can find that in the case $c\le 0$
the potential (\ref{Vk=2}) has exactly the form of
the potential of the linear supersymmetry with the
superpotential
$$
 {\cal W}(x)=f(x)-\frac{f'(x)-2\sqrt{-c}}{2f(x)}.
$$
Here and in what follows we put $\hbar=1$.
If the superpotential ${\cal W}(x)$ is given,
we can consider this relation as differential equation.
Solving this equation,  we obtain the  one-parametric
solution for $k=2$ superpotential $f(x)=W(x)$. Though the
initial potential is the same for linear ($k=1$) and $k=2$
supersymmetric systems, the corresponding
superpartners are different: $V_2^{k=1}=V+{\cal W}'$,
while
$V_2^{k=2}=V+2f'$. Therefore,  in the case $k=2$
one can construct  the one-parametric family of isospectral
potentials as superpartner to $V(x)$.

We will illustrate this by the example of the
infinite square potential well. Without loss of
generality, we can assume that the width of the
potential is equal to 1. This problem is equivalent
to the equation
\[
 \psi''(x)+(2E-\pi^2)\psi(x)=0
\]
with the boundary conditions
\[
 \psi(0)=\psi(1)=0,
\]
and  we can think that $x$ runs over the
segment $[0,\;1]$ only.
The eigenfunctions have the form
\begin{equation}
 \psi_n(x)=\sqrt 2\sin(n+1)x \label{deepES}
\end{equation}
with the energy
\begin{equation}
 E_n=\frac{\pi^2}2n(n+2),    \label{deepEn}
\end{equation}
where the energy of the ground state has been  chosen
to be equal to 0.

Let us consider the infinite square potential well as
a bosonic potential of the  system with the $k=2$ SUSY.
In order to obtain the superpartner of the potential,
we have to find the superpotential $f(x)$. For
simplicity we put $c=0$.  Then to find the
superpotential, we have to solve the  equation
\[
 f(x)-\frac{f'(x)}{2f(x)}=-\frac{\psi _0'(x)}
 {\psi _0(x)}.
\]
The general solution has the form
\begin{equation}\label{fdeep}
 f(x)=\frac{2\pi\sin^2\pi x}{c_0-2\pi x+\sin 2\pi x}.
\end{equation}
In order the superpotential to be well defined
function on $(0,\,1)$, we have to assume that the real
constant $c_0$ can take any value in $\mathbb R$
except  the interval $(0,\,2\pi)$. Let us  note that
the superpartner in the framework of the usual ($k=1$)
supersymmetry is proportional to $\mathop{\rm cosec}^2
\pi x$ and has a pure trigonometric nature,
while this is not the case for
 the $k=2$ SUSY superpartner.

The superpartner of the potential is defined
as $V_f(x)=V(x)+2f'(x)$, and in this case it acquires
the form
\begin{equation}
 V_f(x)={}-\frac{\pi^2}2+16\pi^2\frac{\sin\pi x
 \left(\sin\pi x-\left(\pi x-\frac 12c_0\right)
 \cos\pi x\right)}{\left(c_0-2\pi x+\sin 2\pi x
 \right)^2}. \label{deepVf}
\end{equation}
Acting by  the supercharge $Q^+$ of the form (\ref{Qd2})
with the superpotential (\ref{fdeep}) on the wave
functions (\ref{deepES}) of the square potential well,
we obtain the eigenfunctions of the potential
(\ref{deepVf}),
\[
 \psi _n^{(f)}(x)={\cal N}\left(\frac{4\left(1+n\right)\sin\pi x
 \sin n\pi x}{c_0-2\pi x+\sin 2\pi x}+\left(n^2+\frac
 {2n\left(c_0-2\pi x\right)}{c_0-2\pi x+\sin 2\pi x}
 \right)\sin(1+n)\pi x\right),
\]
given here up  to a normalization constant ${\cal N}$.
One can verify
that $\psi _0^{(f)}(x)$ is identically equal to zero.

\begin{figure}
 \begin{center}
 \epsfxsize=6cm
 \epsfbox{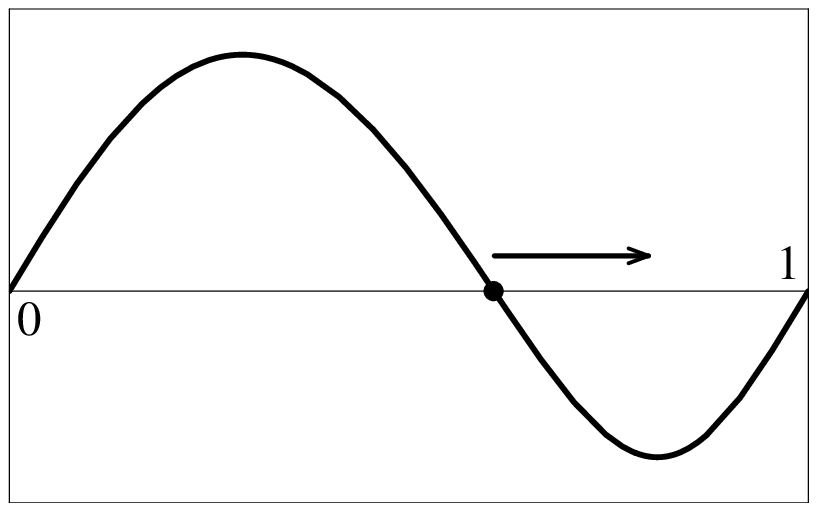}
 \hskip 2cm
 \epsfxsize=6cm
 \epsfbox{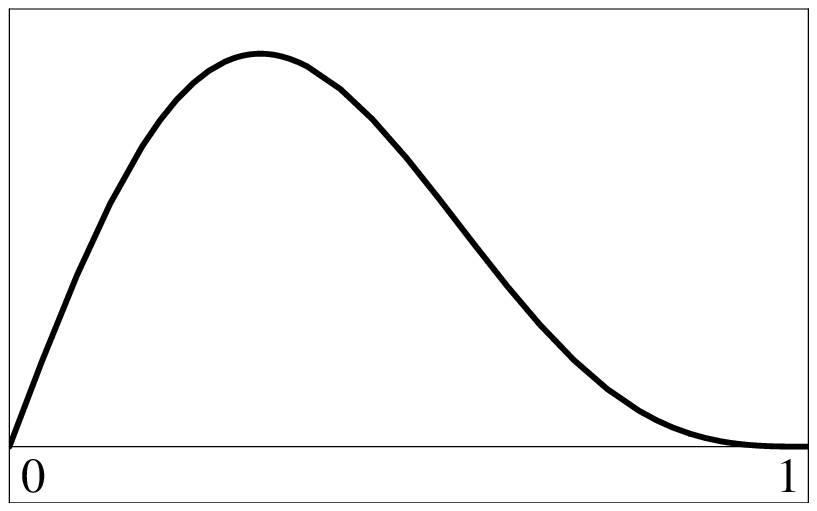}
 \end{center}
 \caption{The plots of the function $\psi _1^{(f)}(x)$
 for the cases $c_0>2\pi$ and $c_0=2\pi$.}
 \label{pic34}
\end{figure}

It is worth noting that the function $\psi _n^{(f)}(x)$
has $n$ nodes when $ c_0\neq 0,2\pi $ and $n-1$ nodes
when $c_0=0$ or $c_0=2 \pi$. This means that in the
latter case $\psi _1^{(f)}(x)$ is the ground state of
the potential (\ref{deepVf}), but this is not valid for
the former case.
\begin{figure}
 \begin{center}
 \epsfxsize=6cm
 \epsfbox{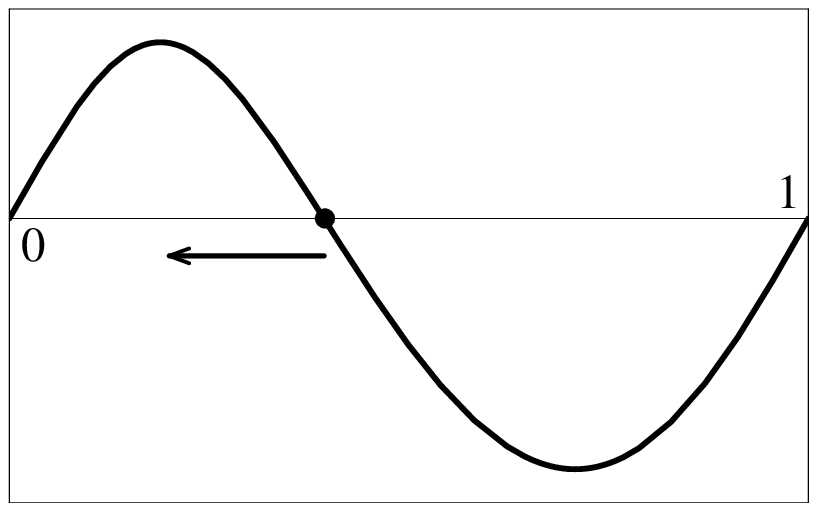}
 \hskip 2cm
 \epsfxsize=6cm
 \epsfbox{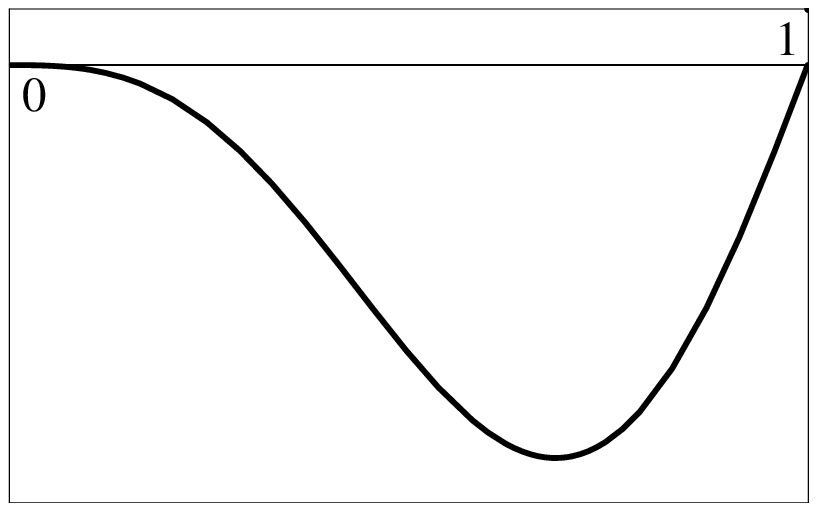}
 \end{center}
 \caption{The plots of the function $\psi _1^{(f)}(x)$
 for the cases $c_0<0$ and $c_0=0$.}
\label{pic12}
\end{figure}
This situation is illustrated on Figs.~\ref{pic12},
\ref{pic34} by the example of the function $\psi _1
^{(f)}(x)$. The arrows indicate direction of the node
motion when $c_0$ goes to $0$ (Fig.~\ref{pic12}) or to
$2\pi$ (Fig.~\ref{pic34}).
For the case $c_0\neq 0,2\pi$, the ground state can be
found by calculating the limit
$\lim_{n\to 0}\psi_n^{(f)}(x)/n$. Having done this,
one can verify that the function obtained in such a way,
\[
 \psi_0(x)=\sqrt{2c_0\left(c_0-2\pi\right)}
 \frac{\sin\pi x}{c_0-2\pi x+\sin 2\pi x},\qquad
 \left\|\psi_0\right\|=1,
\]
is indeed the ground state of the potential. The
function vanishes when $c_0=0$ or $c_0=2\pi$ that
corresponds to the statement above. Therefore,
by a choice of the parameter $c_0$ we can obtain the
unbroken $k=2$ SUSY of the two types, namely, with one
singlet  for $c_0=0$ or $c_0=2\pi$, and with two
singlets with equal energies for other admissible
values of $c_0$.

Thus,  we have illustrated the special case of the
$k=2$ supersymmetric system by the example of the infinite
square
potential well. We have also shown that in the
case $c=0$, the two types of the spectra
represented on Fig.~\ref
{k2sp}${\bf b}$ and  ${\bf c}$ can be realized. The
spectrum of the
first type has exactly the form of that of a system
with the usual supersymmetry in the exact phase. The
spectrum of the second type has rather unusual
properties. It has the form characteristic to a system
with $k=1$ broken supersymmetry (with coinciding two
lowest energy levels), but here the two lowest states
with equal energies are the true supersymmetric singlets
and, therefore, the $k=2$ supersymmetry is unbroken.

\section{Discussion and outlook}
\label{disc}

Let us summarize briefly the obtained results and then
discuss some open problems that deserve further attention.

\begin{itemize}

\item Classical supersymmetry is characterized by the
Poisson algebra being  nonlinear in the Hamiltonian,
and includes the  usual linear supersymmetry as a
particular case.

\item Any supersymmetric system is symplectomorphic to
the supersymmetric system of the canonical form with  the
holomorphic supercharges.

\item The canonical  supersymmetric systems are
separated into the three classes defined by  the behaviour of
the superpotential. In the first class the parameter
$\alpha$ characterizing the order of the superalgebra is
subject to a classical quantization:
$\alpha=k\in \mathbb Z_+$;
in the second class it can take any
non-negative value: $\alpha\in \mathbb R_+$;
in the systems of the third class the fermion degrees of
freedom can be decoupled completely by a canonical
transformation and, hence, their supersymmetry has a rather
fictive nature.

\item We have investigated the nonlinear supersymmetry
with supercharges being polynomial of order $k$ in the
momentum, and for $k=2$ have found  the most general
one-parametric solution of the  Calogero-like form
(\ref{CalogeroH}).
Depending on the value of the parameter $c$,
classically  the Calogero-like $k=2$ supersymmetric system
can be reduced  to the $k=0$, $k=1$
or $k=2$ supersymmetric system of a canonical form.
\end{itemize}

The quantization of a system with nonlinear supersymmetry
is a nontrivial problem due to the quantum anomaly
taking place in general case. We have shown that
\begin{itemize}
\item The anomaly-free quantization of the classical
$k$-supersymmetric system with the holomorphic
supercharges is possible for the superpotential
of the quadratic (\ref{w2}) and exponential forms
(\ref{WExp}) only;

\item The $k$-supersymmetric systems with the
exponential superpotential are closely  related
to the well-known families of the quasi-exactly solvable
systems \cite{shifman};

\item The
$k=2$ supersymmetric Calogero-like systems can be
quantized in the anomaly-free way. The problem of the
quantum anomaly is ``cured"  here by the specific
superpotential-dependent  term of order
$\hbar^2$. Such a quantum term  appeared earlier  in
the operator transformations method of constructing
new solvable potentials from the known ones
\cite{cooper};

\item The general $k=2$ supersymmetry associated with
the Calogero-like systems can be used for producing new
exactly solvable potentials.

\end{itemize}

The supersymmetric systems (\ref{hg0})
given by the superpotentials of the third type
are canonically equivalent to the system
(\ref{decouple}) with the completely decoupled  fermion
degrees of freedom. On the other hand, the $k=1$
quantum analogues of (\ref{hg0}) with
superpotential of the third type are the systems
with spontaneously broken linear supersymmetry.
In this case the nilpotent operators
$\tilde Q{}^\pm=Q^\pm/\sqrt{2H}$
are well defined and their anticommutator
is  equal to $1$, i.e. $\tilde Q{}^\pm$ look like
$k=0$ quantum supercharges.
However, these operators are not
decoupled from the even operators $x$ and $p$.
Therefore, the question is whether
the quantum analogue of the above mentioned
canonical transformations exists.
If so,  the quantum fermion degrees of freedom could be
completely decoupled and the initial $k=1$ supersymmetric
system  would be reduced to the $k=0$ supersymmetric
system in correspondence with the classical picture.
On the other hand,  if such a transformation does not exist
(at least for some superpotentials $W(x)$),
we face a sort of quantum transmutation.
The classical equivalence of the Calogero-like
$k=2$ supersymmetric systems to the
$k=0$ (for $c>0$)  and  to the $k=1$  (for $c<0$)
supersymmetric systems of the canonical form (\ref{hg0})
is also based on the existence of the
corresponding canonical transformations.
If for the quantum Calogero-like $k=2$ supersymmetric
systems (\ref{HD})--(\ref{WD})
the quantum analogues of the above mentioned canonical
transformations do not exist, we again have a quantum
transmutation.

The supersymmetric system
(\ref{hg0}) represents the whole class of the
symplectomorphic systems (\ref{h1}).
However, the quantization apparently breaks
the equivalence of the systems (\ref{h1}) with
different functions $M(W^2)$. Therefore the
quantization of the systems (\ref{h1})
may lead to different nontrivial quantum
systems with $k$-supersymmetry. Most likely,
the anomaly-free quantization could be possible
only  for some special cases of the function
$M(W^2)$ and the superpotential $W(x)$.
In this context it is necessary to stress  that the
quantum anomaly is specific not only for the nonlinear
supersymmetry. The quantization of  the general system
(\ref{h1}) with $k=1$ and $M(W^2)\not\equiv 0$
gives rise to the quantum anomaly. {}From this
view-point one can say that the system of the form
(\ref{hg0}) underlies  the quantum linear supersymmetry
since the quantization of such a system with arbitrary
superpotential does not reveal any anomaly. However,
it would be interesting to find at least one example of
the quantum-mechanical system of the non-standard form
(\ref{h1}) with the linear supersymmetry. Since for the
nontrivial cases the
quantum analogue of the transformation (\ref{CT})
is non-local,  the corresponding system has to operate with
non-local objects. It is possible that the supersymmetry
of the pure parabosonic systems
\cite{susy-pb} can be
arrived at in this way.
We hope that further investigations will shed light
on the relation between the classical and quantum
supersymmetries in the context  on the described
hypothetical quantum transmutations and the quantum
anomaly problem.

Though we have considered the
concept of the nonlinear supersymmetry
for one-dimensional  systems,
it can be extended to the higher dimensional
systems as well. For example, the two-dimensional system
describing the motion of a charged particle in external
magnetic field admits the  nonlinear supersymmetries.
The Pauli Hamiltonian of
such a  system is given by ($\hbar=m=e=1$)
$H=\frac{1}{2}({\cal P}_1^2+{\cal P}_2^2
+g\varepsilon_{ij}\partial_iA_jN)$, where
${\cal P}_i={}-i\partial_i+A_i$, and $A_i$, $i=1,2$,
form the 2D vector gauge potential.
For the gyromagnetic ratio $g=2$ this
system reveals the usual linear supersymmetry both
at the classical and quantum levels \cite{cooper}.
It turns out that for  $g=2k$ the system
possesses  the {$k$-supersymmetry}.
To show this let us
introduce the complex variables
${\tilde z{}^\pm={{\cal P}_1\pm i{\cal P}_2}}$.
In terms of these variables the Hamiltonian can
be rewritten as
$H=\frac{1}{2}(\tilde z{}^+\tilde z{}^-
+ik\{\tilde z{}^+,\tilde z{}^-
\}N)$.
But this is exactly the form of the Hamiltonian
(\ref{hg0}) with the variables $z$, $\bar{z}$ changed for
$\tilde{z}{}^+$, $\tilde{z}{}^-$.
Therefore, the Hamiltonian commutes with the
supercharges $Q^\pm=(\tilde z{}^\mp)^k\theta^\pm$
and, as a consequence, the system possesses the
classical $k$-supersymmetry. The
quantization of this system reveals the properties similar
to those for  the $k$-supersymmetric  system with the
holomorphic
supercharges:  the requirement of
conservation of the nonlinear supersymmetry
leads to the
restrictions on the form of the magnetic field
$B=\varepsilon_{ij}\partial_iA_j$.
One can show that as in  the
one-dimensional case,
the form of the magnetic field  turns out to be
restricted by the quadratic case
and by the  sum of exponential functions in the variables
$x_i$ \cite{inprog}.
We suppose that such systems are also closely
related to the  two-dimensional quasi-exactly
solvable problems.

To conclude, we  note that our
investigation of the nonlinear supersymmetry was
motivated by the existence of nonlinear supersymmetry
in pure parabosonic systems \cite{susy-pb}, where
the reflection (parity) operator $R$ plays the role
of the $Z_2$-grading operator.
Such a special role of the reflection operator
was the main idea in the construction of the
minimally bosonized $k=1$ supersymmetric quantum
mechanics \cite{rbos,GPZ99}, where the role of the
superpotential  is played by the arbitrary odd function.
Therefore, the construction of the bosonized nonlinear
supersymmetry is an attractive problem  which we hope
to consider elsewhere \cite{inprog}.

When this paper was  finished, the interesting papers
\cite{Aoyama1,Aoyama2} devoted to the
development of the $1D$  nonlinear supersymmetry
of ref. \cite{AoyamaNP} have appeared.
We note that the generalization of the
$k$-supersymmetry with
the holomorphic supercharges found in ref. \cite{Aoyama2}
and possessing  the typical Calogero-like structure of
the  $k=2$ supersymmetry from  Section 5
could also be treated in  the  context of the quasi-exactly
solvable systems discussed here.

\vskip0.5cm
{\bf Acknowledgements}
\vskip5mm

The work was supported  by the
grants 1980619, 1010073 and 3000006 from
FONDECYT (Chile)
and by DYCIT (USACH).

\appendix
\section*{Appendix}

Here we show that
the arbitrariness in the definition of the superpotential,
$
 \tilde W^2(x)=W^2(x)+const,
$
can not be used to change the type of the classical
supersymmetry of the given system, i.e. we  demonstrate
the invariance of the  classification obtained in  Section
\ref{c2}.
For the purpose,  it is enough to prove that the redefinition
of the superpotential can not  provoke a transition
between the adjacent classes of supersymmetry.

\begin{figure}[ht]
 \begin{center}
 \epsfxsize=8cm
 \epsfbox{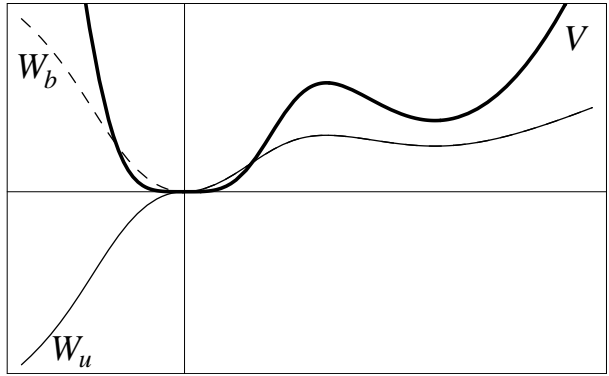}
 \end{center}
 \caption{The plots of a potential $V(x)$ and the
 corresponding superpotentials $W_u(x)$ and $W_b(x)$.}
 \label{pic}
\end{figure}

Let us start by analysing  the possibility of the transition
from the supersymmetry of the first type with the
superpotential unbounded from  below to the
supersymmetry of the second type.
For  the sake of simplicity we assume that the
potential $V(x)$ has finite number of minima,
the lowest one is at the origin and is  equal to zero.
For such a potential,
the superpotentials of the first and second types obey the
relation $W^2(x)=V(x)$.
Suppose that  the bounded and unbounded
superpotentials $W_u(x)$ and $W_b(x)$ with the
associated supersymmetries  of the first and second
types correspond to the potential.
The superpotential $W_b(x)$ is a continuously
differentiable function at $x=0$ only when the potential has
the local behaviour $V(x)=x^4+O(x^5)$. Indeed, if
$V(x)=x^2+O(x^3)$, the bounded superpotential locally
is $W_b(x)=|x|+O(x^2)$, and as a
consequence, its derivative is not regular at the origin.
The case of the potential admitting the regular superpotential
$W_b(x)$ is presented  on the Fig.~\ref{pic}.
In this case the bounded and unbounded superpotentials
$W_b(x)$ and $W_u(x)$ coincide for positive values of
$x$ and have different sign for negative $x$:
$
 W_u(x)=\mathop{\rm sign}(x)W_b(x).
$
If we have the nonlinear supersymmetry of the first
type, i.e. $L(x)=kW_u'(x)$, $k\in\mathbb N$, then
rewriting this function in terms of $W_b$ we obtain
$L(x)=k\mathop{\rm sign}(x)W_b'(x)$. The
function $L(x)$ is not good  in terms of $W_b$ in the
sense that there exist no canonical transformation
that could reduce this system to that with the
supersymmetry of the second type given by Eq. (\ref{h3}).
The potential with several minima can be considered in a
similar way.

The case of supersymmetry of the first type with a
bounded superpotential merits the special analysis.
After the transition (\ref{trans}) to the
superpotential $\tilde W$ of the second type, the
function $L(x)$ can be rewritten as
$L(x)=k\tilde W'\tilde W/(\tilde W^2-\tilde w)^{1/2}$,
$\tilde w=const$.
The function $(\tilde W^2-\tilde w)^{-1/2}$
has singularities at the points where the initial superpotential
vanishes. Therefore,  the type of the supersymmetry can
not be changed either.

Now let us consider the  possibility of transition
from the supersymmetry of the second type to that of
the third type by means of the change (\ref{trans}).
We suppose that the superpotential
$W(x)$ of a given system has a minimum equal to zero at
the origin of coordinates. The superpotential of the
third type defined by $W^2(x)=\tilde W^2(x)-w$ has the
minimum equal to $w$ at the origin. Rewriting $L(x)$
in terms of the new superpotential, we obtain
$L(x)=\alpha W'=\alpha\tilde W'\tilde W/(\tilde
W^2-w)
^{1/2}$. There exists no canonical transformation
(\ref{CT}) that could remove the corresponding nilpotent
part of the Hamiltonian since the function
$\tilde W/(\tilde W^2-w)^{1/2}$ is singular at the
origin. The case of the  superpotential with several  minima
can be treated in a similar way.
This completes our proof of the invariance of the obtained
classification of the classical supersymmetries.

\end{document}